\documentclass[%
reprint,
superscriptaddress,
 amsmath,amssymb,
 aps,
]{revtex4-1}
\pdfoutput=1
\usepackage{graphicx}
\usepackage{dcolumn}
\usepackage{bm}
\usepackage{cancel}
\usepackage{color}
\usepackage{soul} 
\usepackage{tabularx}
\newcolumntype{L}{>{\raggedright\arraybackslash}X}
\begin{document}

\makeatletter
\newcommand{\bottomrule}{\unskip\lrstrut\\\noalign{\hline@rule}{}}%
\newcommand{\midrule}{\unskip\lrstrut\\\noalign{\hline@rule}{}}%
\makeatother
\newcommand{\PFS}[1]{\begingroup\color{green}#1\endgroup}
\newcommand{\TODO}[1]{\begingroup\color{red}\emph{\bfseries{#1}}\endgroup}
\newcommand{\JJ}[1]{\begingroup\color{red}\emph{\bfseries{#1}}\endgroup}
\newcommand{\chem}[1]{\texttt{#1}}


\title{Forman-Ricci Curvature for Hypergraphs}

\author{Wilmer Leal}
 \altaffiliation[]{wilmer@bioinf.uni-leipzig.de
}
\affiliation{Bioinformatics Group, Department of Computer Science,
  Universit{\"a}t Leipzig, H\"artelstra{\ss}e 16-18, 04107 Leipzig, Germany}
\affiliation{Max Planck Institute for Mathematics in the Sciences, Inselstra{\ss}e 22, 04103 Leipzig, Germany}

\author{Guillermo Restrepo}%
\affiliation{Max Planck Institute for Mathematics in the Sciences, Inselstra{\ss}e 22, 04103 Leipzig, Germany}
\affiliation{Interdisciplinary Center for Bioinformatics, Universit{\"a}t
  Leipzig, H{\"a}rtelstra{\ss}e 16-18, 04107 Leipzig, Germany}
  
\author{Peter F. Stadler}%
\affiliation{Bioinformatics Group, Department of Computer Science,
  Universit{\"a}t Leipzig, H\"artelstra{\ss}e 16-18, 04107 Leipzig, Germany}
\affiliation{Max Planck Institute for Mathematics in the Sciences,
  Inselstra{\ss}e 22, 04103 Leipzig, Germany}
  \affiliation{Interdisciplinary Center for Bioinformatics, Universit{\"a}t
  Leipzig, H{\"a}rtelstra{\ss}e 16-18, 04107 Leipzig, Germany}
  \affiliation{Institute for Theoretical Chemistry, University of Vienna,
  W{\"a}hringerstra{\ss}e 17, 1090 Vienna, Austria}
   \affiliation{Facultad de Ciencias, Universidad Nacional de Colombia, KR 30-45 3, 111321, Bogot\'a, Colombia}
  \affiliation{The Santa Fe Institute, 1399 Hyde Park Rd., 87501, Santa Fe, New Mexico, USA}

 \author{J\"urgen Jost}%
\affiliation{Max Planck Institute for Mathematics in the Sciences,
  Inselstra{\ss}e 22, 04103 Leipzig, Germany}
  \affiliation{The Santa Fe Institute, 1399 Hyde Park Rd., 87501, Santa Fe, New Mexico, USA}




\date{\today}

\begin{abstract}
In contrast to graph-based models for complex networks, hypergraphs are more general structures going beyond binary relations of graphs.  For graphs, statistics gauging different aspects of their structures have been devised and there is undergoing research for devising them for hypergraphs.  Forman-Ricci curvature is a statistics for graphs, which is based on Riemannian geometry, and that stresses the relational character of vertices in a network through the analysis of edges rather than vertices.  In spite of the different applications of this curvature, it has not yet been formulated for hypergraphs.  Here we devise the Forman-Ricci curvature for directed and undirected hypergraphs, where the curvature for graphs is a particular case.  We report its upper and lower bounds and the respective bounds for the graph case.  The curvature quantifies the trade-off between hyperedge(arc) size and the degree of participation of hyperedge(arc) vertices in other hyperedges(arcs).  We calculated the curvature for two large networks: Wikipedia vote network and \emph{Escherichia coli} metabolic network.  In the first case the curvature is ruled by hyperedge size, while in the second by hyperedge degree.  We found that the number of users involved in Wikipedia elections goes hand-in-hand with the participation of experienced users.  The curvature values of the metabolic network allowed detecting redundant and bottle neck reactions.  It is found that ADP phosphorilation is the metabolic bottle neck reaction but that the reverse reaction is not that central for the metabolism.

\end{abstract}

\pacs{Valid PACS appear here}
\maketitle


\section{\label{sec:level1}Introduction}
Hypergraphs are used to model systems whose objects have not only binary
relationships; instead, interactions simultaneously involve multiple
members \cite{Bretto, Jost-MATCH}.  Examples of these systems are found in
physics, biology, chemistry, computer science, combinatorial optimization,
scientometrics and several other fields \cite{Klamt, Gallo, Bretto,
  Barbosa2018, Michoel2012, Vazquez2008, Xiong2018}. Hypergraphs
  reduce to (ordinary) graphs when all relationships (hyperedges) are
  binary.  Graphs have been widely used as a mathematical model for
different systems and their mathematical properties have been extensively
studied, which include devising statistics gauging aspects of their
structures, such as vertex degree and its distributions, clustering
coefficients, betweenness centrality and more recently Forman-Ricci
curvature.

As hypergraphs are a generalization of graphs, several of the graph
statistics have been extended to hypergraphs, e.g. vertex and hyperedge
degrees, clustering coefficients \cite{Klamt, Estrada} and spectral
properties \cite{Zhou}. Most of the commonly used quantities
  focus on vertices.  As the crucial structure of a graph is, however,
given by the set of its edges rather than by its vertices, we should
systematically define and evaluate quantities assigned to the edges rather
than to the vertices. In this paper we develop the Forman-Ricci curvature
for hypergraphs (directed and undirected) and calculate it for networks of
different sizes and research fields.

\section{Forman-Ricci curvature of edges/arcs in graphs}

Recently various notions of ``curvature'' have been proposed for graphs and
other, more general, discrete structures and applied to detect various
local or global properties of such structures \cite{2017arXiv171207600S, 2017arXiv170700180W, 74177, 2017CSF10150S,
    2016arXiv160708654W, 2016arXiv160807838W, 2016arXiv160504662S,
    2016arXiv160406634W, 2016JSMTE063206S, Jost-MATCH}. The name of ``curvature''
may seem somewhat strange in this context. In differential, and more
abstractly, in Riemannian geometry, curvature has been found to encode and
express local and global features of smooth manifolds equipped with metric
tensors \cite{Bauer2017}. Those features themselves usually do not depend on an underlying
smooth structure, and this has lead to abstract theories of generalized
curvatures on metric spaces. On graphs, these generalized curvatures are
particularly easy to define and to evaluate. They can also shed
considerable light on other quantities that have been introduced in network
analysis without such a clear conceptual background as those
curvatures. The simplest among these generalized curvatures is the Ricci
curvature introduced by Forman for simplicial complexes \cite{Forman2003}. As graphs are
one-dimensional simplicial complexes, we can readily evaluate this
curvature. As explained in detail in Section \ref{undi}, for an edge
$e=\{i,j\}$ with vertices $i,j$ with degrees $d_i$ and $d_j$ (the degree of a
vertex is the number of its neighbors, that is, of those other vertices
that are directly connected to it by an edge), the Forman-Ricci curvature
is simply $4-d_i-d_j$. The number $4$ serves the purpose of normalization,
to make the curvature of cycle graphs vanish. The minus signs are also
conventional, to align this curvature with the Ricci curvature of
Riemannian geometry. Thus, edges connecting vertices of large degree have
very negative curvature values, and the first step in the analysis of an
empirical network might consist in identifying the most negatively curved
edges as the most important ones for the cohesion of the network or for the
canalization and distribution of information or activity in the
network.

Since the definition of the Forman-Ricci curvature of an edge in an
undirected graph is so clear and simple, it can be readily generalized to,
for instance, directed or weighted graphs, and also to structures in which
more than two elements are related. Forman himself had introduced this
curvature notion already for possibly weighted, simplicial complexes \cite{Forman2003}. A
simplicial complex is characterized by the requirement that whenever a
collection of $k$ elements stands in relation, then this also holds for any
subcollection. This leads to mathematically very nice properties, and
simplicial complexes are basic structures in algebraic topology, but for
the modelling and analysis of empirical data sets, we may want to relax or
perhaps even completely abandon that condition. That leads us to
hypergraphs, which are collections of vertices (undirected hypergraphs) or collections endowed with direction (directed hypergraphs).  Examples of the former are elections, where a subset of voters is an election an the collection of elections constitutes the hypergraph. Chemical reactions \cite{Klamt, Stadler-chapter, Fagerberg2018} and particle scatterings are instances of directed hypergraphs, where some starting materials are transformed into some products.  For hypergraphs, in
principle, various generalizations of the Forman-Ricci graph curvature are
possible. It is a main contribution of this paper to identify that notion
of Forman-Ricci curvature for (un)directed hypergraphs that is best adapted to their structure and to investigate its
properties. We also apply this to concrete empirical hypernetworks, a
social and a metabolic one.

In this section we briefly summarize the results of the Forman-Ricci curvature for graphs and then generalize the curvature for hypergraphs.

\subsection{Undirected graphs}
\label{undi}

\begin{figure}[h]                                             
  \centering\includegraphics[width=.35\textwidth]{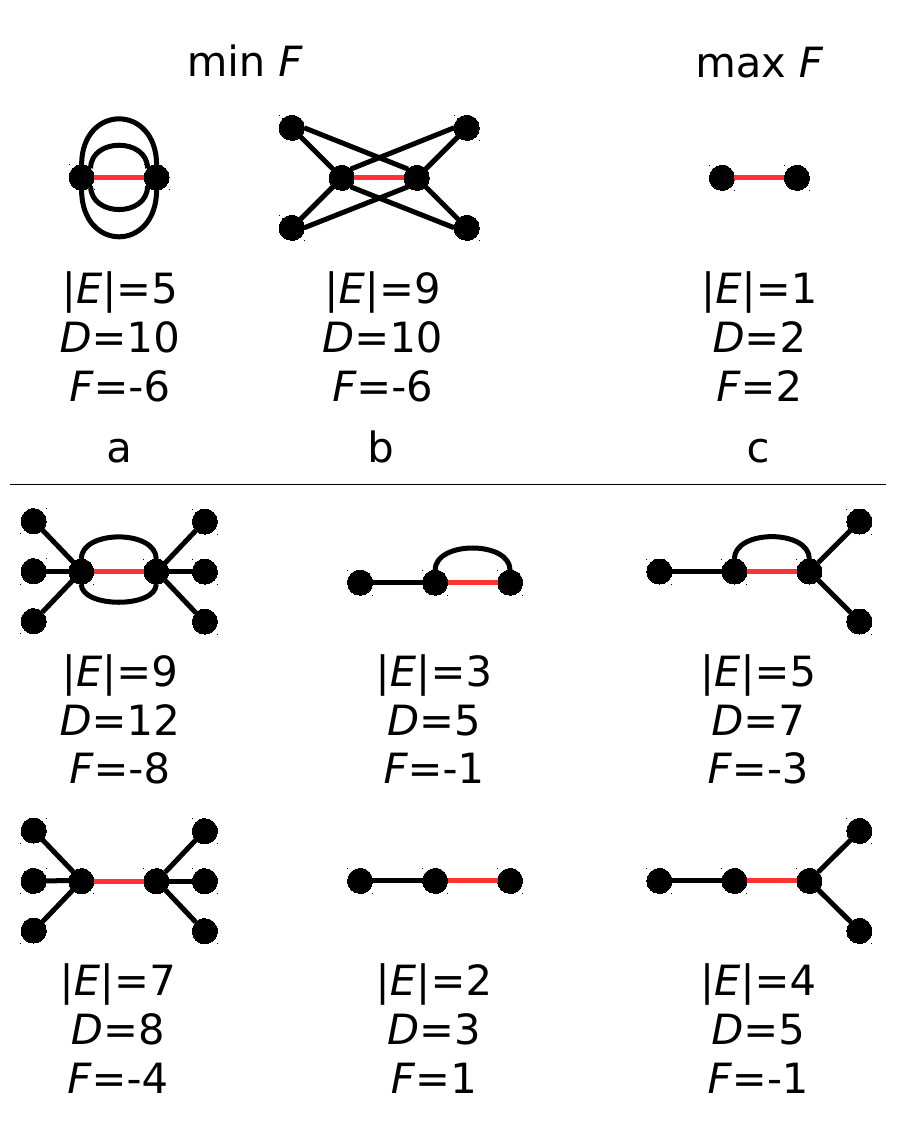}
  \caption{Forman-Ricci curvatures $F(e)$ calculated for the red edge $e$
    of the depicted undirected graphs.}
\label{fr-graphs}
\end{figure}

Let $G=(V,E)$ be a (multi)graph with vertex set $V$ and multiset of edges
$E$.  The Forman-Ricci curvature of an edge $e=\{i,j\}\in E$, as introduced
in \cite{2017arXiv171207600S}, is given by:
\begin{equation}
  F(e)=w_e\Bigg(\frac{w_i}{w_e} + \frac{w_j}{w_e} -
  \sum_{e_l \sim i} \frac{w_i}{\sqrt[]{w_e w_{e_l}}} -
  \sum_{e_l \sim j} \frac{w_j}{\sqrt[]{w_e w_{e_l}}}\Bigg)
\label{FR-curvature}
\end{equation}
where $w_e$ denotes the weight of the edge $e$, $w_i$ and $w_j$ are the
weights of vertices $i$ and $j$, respectively. The sums over $e_l\sim k$
run over all edges $e_l$ incident on the vertex $k$ excluding $e$.  The
curvature for the unweighted multigraph, with vertex and edge weights set to 1, is given by \cite{Jost-MATCH}
\begin{equation}
F(e)=4-d_i-d_j 
\label{curvature-sg}
\end{equation}
where $d_k$ is the vertex degree of $k$. Defining $D=\sum_{k\in e}{d_k}$ we have 
\begin{equation}
F(e)=4-D
\label{curvature-sgD}
\end{equation}
As a multigraph may have repeated edges, whose number is independent of the
number of vertices, the bounds for $F(e)$ shall be expressed as a function
of the known number of edges, namely, $|E|$.  Therefore,
$2(2-|E|)\leq F(e)\leq 2$.  The lower bound is attained when $d_k=|E|$ for
every $k\in e$, therefore $D=2|E|$ (Figure \ref{fr-graphs}a).  In turn
$F(e)=2$, for an isolated edge $e$ (Figure \ref{fr-graphs}c).  In contrast
to the multigraph case, for simple unweighted graphs, the lower bound can
be expressed as a function of the number of vertices: $2(3-|V|)\leq F(e)$,
which is obtained for $d_k=|V|-1$ for every $k\in e$, i.e., $D=2(|V|-1)$
(Figure \ref{fr-graphs}b).  As for multigraphs, $F(e)$ reaches its maximum
value ($F(e)= 2$) for an isolated edge (Figure \ref{fr-graphs}c).

As seen in Figure \ref{fr-graphs}, Forman-Ricci curvature quantifies the
degree of spread of the vertices in $e$, from maximum spread
(corresponding to $\min F(e)$) to minimum spread (attained when $\max F(e)$).

\subsection{Directed graphs}

\begin{figure}[h]                                             
\centering                                                    \includegraphics[width=.45\textwidth]{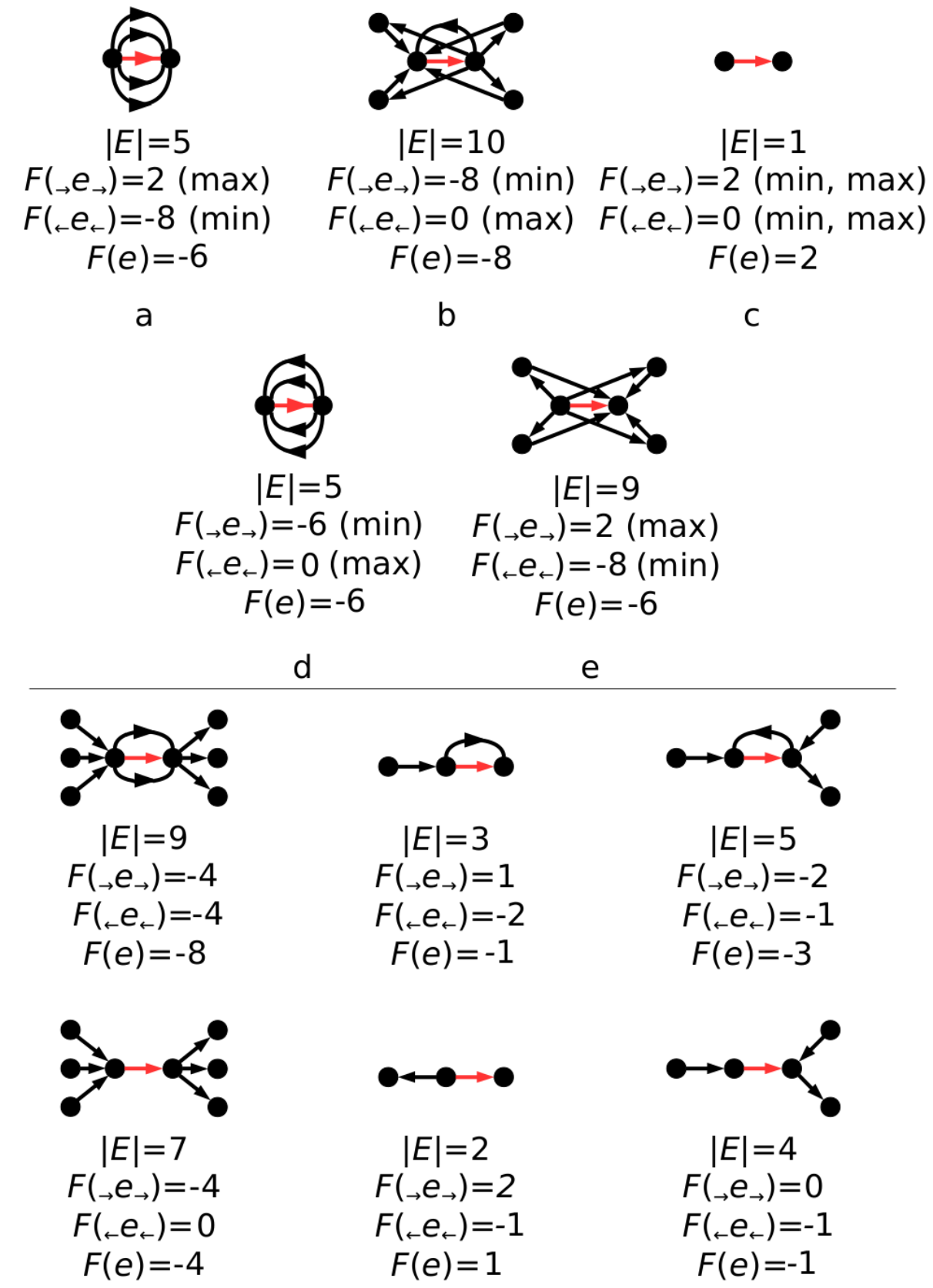}
\caption{Forman-Ricci curvatures $F(_\rightarrow e_\rightarrow)$, $F(_\leftarrow e_\leftarrow)$, and $F(e)$ calculated for the red arc $e$ of the depicted directed graphs.}
\label{fr-directed-graphs}
\end{figure}

Here we are interested in an unweighted directed multigraph $G=(V,E)$,
where $e=(i,j)\in E$ is an \emph{arc} (directed edge), and $i,j\in V$.
Equation \ref{curvature-sg} indicates that the curvature of an edge depends
on the degree of its vertices.  As in a simple directed graph the
degree can be split into in- and out-degree. The curvature of $e=(i,j)$ is
defined in terms of in- and out-degrees as well
\cite{2016arXiv160504662S}. There are different possibilities for the
realization of the curvature, depending on the meaning one assigns to
it. Here we emphasize the directed spread or \emph{flow through} $e$, i.e.,
following the direction of the arc.  Therefore, we consider the incoming
arcs on $i$ (in-degree of $i$, $\text{in}(i)$) and the outgoing arcs from $j$ (out-degree of $j$, $\text{out}(j)$).
When we separate the curvature in \eqref{curvature-sg} into the contribution $2-d_i$ of $i$ and $2-d_j$ of $j$ and also note that the edge $e$ counts for the degrees of $i$ and $j$, but neither for the in-degree of $i$ nor for the out-degree of $j$, then a curvature accounting for the in-flow at $i$ ($F(_\rightarrow e)$)
and another for the out-flow at $j$ ($F(e_\rightarrow)$) is defined as
\begin{equation}
\begin{split}
F(_\rightarrow e)&=  1-\text{in}(i)\\
F(e_\rightarrow)&= 1-\text{out}(j)\,.
\end{split}
\label{in-out-directed-graph}
\end{equation}
Both are bounded below by $2-|E|$ for $\text{in}(i)=\text{out}(j)=|E|-1$,
and bounded above by $1$ when $\text{in}(i)=\text{out}(j)=0$ (Figure \ref{fr-directed-graphs}a).  For the
simple directed graph the lower bound for both, in- and out-flow, is
$2-|V|$, for $\text{in}(i)=\text{out}(j)=|V|-1$ (Figure
\ref{fr-directed-graphs}b).  The upper bound is reached, in both cases, when
$\text{in}(i)=\text{out}(j)=0$ (Figure \ref{fr-directed-graphs}c).  The
curvature accounting for the flow through $e=(i,j)$ is then given by
\begin{equation}
\begin{split}
F(_\rightarrow e_\rightarrow) & =F(_\rightarrow e)+F(e_\rightarrow)\\
 & =2-\text{in}(i)-\text{out}(j)
\end{split}
\label{directed-graph}
\end{equation}
where $2(2-|E|)\leq F(_\rightarrow e_\rightarrow) \leq 2$ for the
multigraph case and $2(2-|V|)\leq F(_\rightarrow e_\rightarrow) \leq 2$ in
the simple graph case.  Figure \ref{fr-directed-graphs}c shows the case
where $F(_\rightarrow e_\rightarrow)=2$.  Some further examples of
calculations of curvatures $F(_\rightarrow e_\rightarrow)$ are shown in
Figure \ref{fr-directed-graphs}.

If the \emph{flow-loss} along $e$ is to be considered, two additional
curvatures are calculated that account for the flow loss at $i$
($F(_\leftarrow e)$) and at $j$ ($F(e_\leftarrow)$).  Thus
\begin{equation}
\begin{split}
F(_\leftarrow e)&=1-\text{out}(i)\\
F(e_\leftarrow)&=1-\text{in}(j)
\end{split}
\label{in-out-contra-directed-graph}
\end{equation}
both bounded below by $1-|E|$, for $\text{out}(i)=\text{in}(j)=|E|$, and
bounded above by $0$ for $\text{out}(i)=\text{in}(j)=1$ (Figure \ref{fr-directed-graphs}d).  For the simple
directed graph we have $2-|V|\leq F(_\leftarrow e)\leq 0$ and
$2-|V|\leq F(e_\leftarrow)\leq 0$.
Hence, the curvature for the flow-loss along $e=(i,j)$ is
\begin{equation}
\begin{split}
F(_\leftarrow e_\leftarrow) & =F(_\leftarrow e)+F(e_\leftarrow)\\
 & =2-\text{out}(i)-\text{in}(j)
\end{split}
\label{contra-directed-graph}
\end{equation}
where $2(1-|E|)\leq F(_\leftarrow e_\leftarrow) \leq 0$ (Figures
\ref{fr-directed-graphs}a-e) holds in the multigraph case and
$2(2-|V|)\leq F(_\rightarrow e_\rightarrow) \leq 0$ in the simple graph
case. Some further examples are shown in Figure \ref{fr-directed-graphs}.

A curvature accounting for the \emph{total flow over} $e$ is then computed as
\begin{equation}
F(e) =F(_\rightarrow e_\rightarrow)+F(_\leftarrow e_\leftarrow)\\
\label{total-directed-graph}
\end{equation}

In the following section we extend the Forman-Ricci curvature to hypergraphs.

\section{Forman-Ricci curvature of hyper(edges/arcs) in hypergraphs}

Given a set of vertices $V$, a graph is a collection of subsets (edges) of $V$, all of which comprise only two elements.  If we call the cardinality of each subset  its \emph{size}, then a graph is a collection of subsets of size two.   In a hypergraph, the size of the  subsets is no longer restricted, and subsets of any size are allowed.

\subsection{Undirected hypergraphs}

\begin{figure}[h]                                             
  \centering
  \includegraphics[width=.5\textwidth]{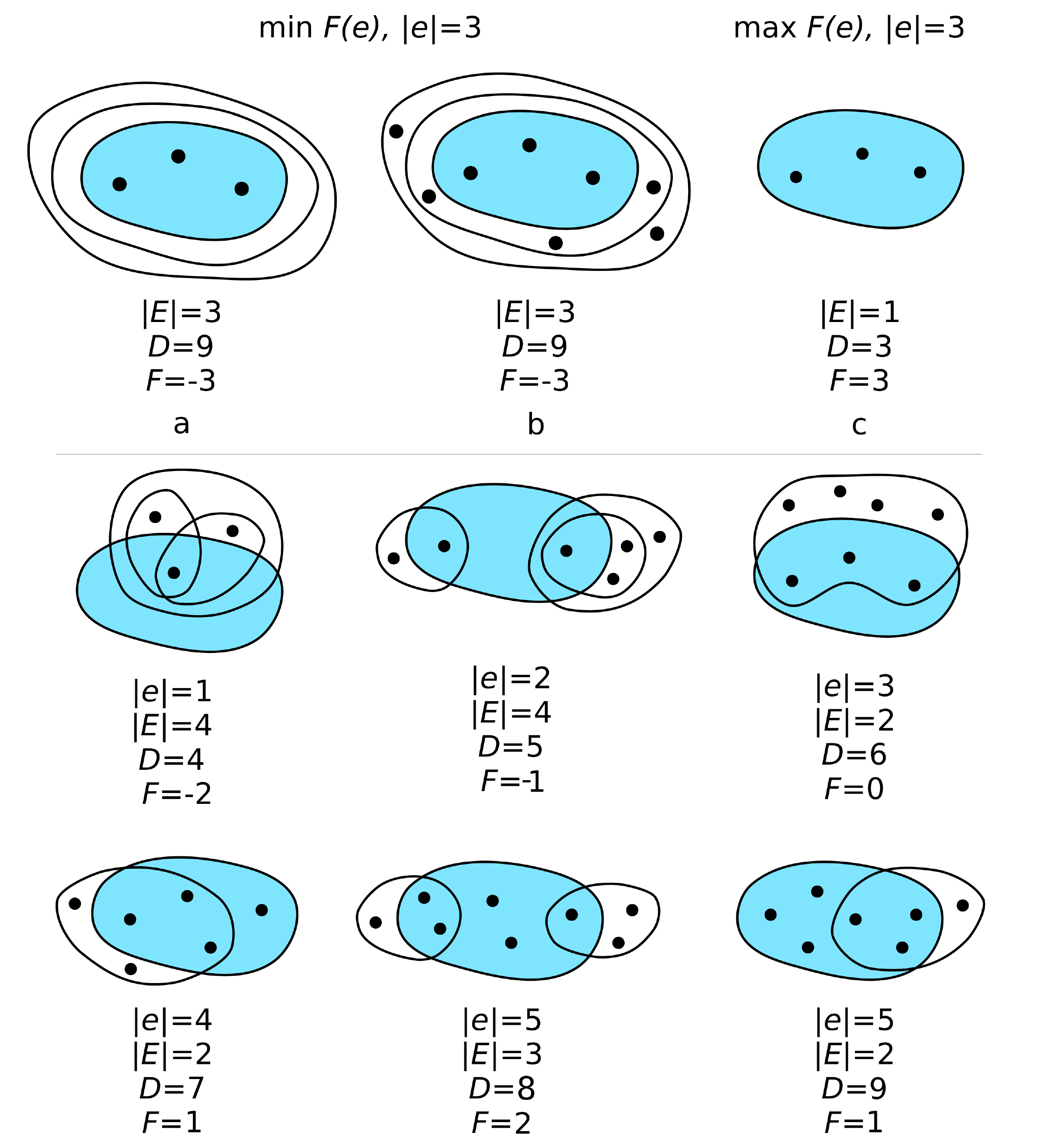}
  \caption{Forman-Ricci curvatures $F(e)$ calculated for the blue hyperedge
    $e$ of the depicted hypergraphs.}
\label{fr-hypergraphs}
\end{figure}

An undirected \emph{hypergraph} $H=(V,E)$ consists of a set $V$ of vertices
and a multiset $E$ of subsets of $V$, called \emph{hyperedges}, such that $e\subseteq V$, i.e. $|e|\leq |V|$, for $e\in E$. Some examples of hypergraphs are
shown in Figure \ref{fr-hypergraphs}.

Separating the contributions of vertices $i$ and $j$ in Equation
\ref{FR-curvature}, it can be rewritten as:
\begin{equation}
  F(e) =w_e\Bigg[\Bigg(\frac{w_i}{w_e} - \sum_{e_l \sim i} \frac{w_i}{\sqrt[]{w_e w_{e_l}}}\Bigg) + \Bigg(\frac{w_j}{w_e} - \sum_{e_l \sim j} \frac{w_j}{\sqrt[]{w_e w_{e_l}}}\Bigg)\Bigg]
\end{equation}
furthermore,
\begin{equation}
  F(e) =w_e\Bigg[\sum_{k \in e}\Bigg(\frac{w_k}{w_e} - \sum_{e_l \sim k} \frac{w_k}{\sqrt[]{w_e w_{e_l}}}\Bigg)\Bigg]
\label{FR-hyper-curvature}
\end{equation}
Since Equation \ref{FR-hyper-curvature} no longer restricts $e$ to size
two, we present it as the Forman-Ricci curvature of the hyperedge $e$. For
the unweighted hypergraph, where all vertex weights are equal to 1, this
expression simplifies to 
\begin{equation}
  F(e) =\sum_{k \in e}\Bigg(2 - d_k\Bigg)=2|e|-\sum_{k \in e}d_k =2|e|-D
\label{FR-hyper-curvature-unw}
\end{equation}
which is bounded below by $|e|(2-|E|)$ when $d_k=|E|$ for every
$k \in e$, and bounded above by $1$ when $D=|e|$. In other words, the
minimum curvature occurs when every vertex in $e$ belongs to each hyperedge
(Figures \ref{fr-hypergraphs}a,b); the maximum is attained for an isolated
hyperedge (Figure \ref{fr-hypergraphs}c).

For the particular case of simple hypergraphs, we therefore have the lower
bound $2|e|(1-2^{|V|-2})$ when $d_k=|\mathcal{P}(V\setminus\{k\})|$ for
every $k \in e$, and the upper bound $|V|$, when $E=\{V\}$.
Note that in hypergraphs $|e|\leq |V|$, therefore the minimum value $|e|$
may reach $1$, unlike graphs.  In such a case,
$2(1-2^{|V|-2})\leq F(e)\leq |V|$.  Some further examples of curvature for
hypergraphs are shown in Figure \ref{fr-hypergraphs}.

\subsection{Directed hypergraphs}

\begin{figure}[h]                                             
\centering                                                    \includegraphics[width=.45\textwidth]{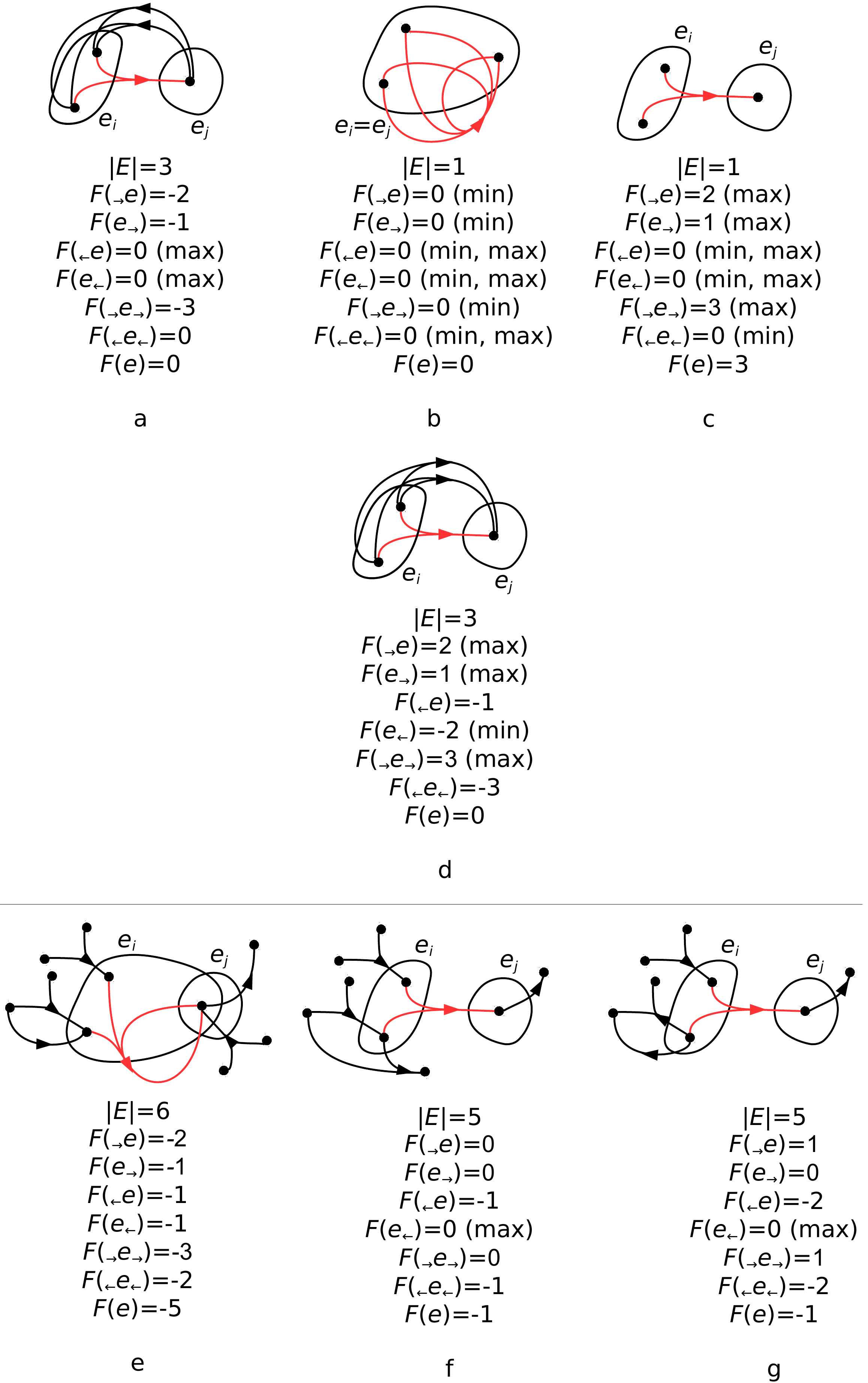}
\caption{Forman-Ricci curvatures $F(_\rightarrow e_\rightarrow)$, $F(_\leftarrow e_\leftarrow)$, and $F(e)$ calculated for the red hyperarc $e$, connecting vertices in $e_i$ with those in $e_j$, of the depicted hypergraphs.}
\label{fr-directed-hypergraphs}
\end{figure}

In a directed hypergraph, each hyperedge is composed of two subsets of
  vertices: the \emph{tail} and the \emph{head} of the hyperedge.
Formally, we say that a \emph{directed hypergraph} $H$ is the couple
$(V,E)$ with $V$ a set of vertices and $E$ a multiset of hyperarcs. A
\emph{hyperarc} is a pair $e=(e_i,e_j)$,
where $e_i\subseteq V$ and $e_j\subseteq V$ are called its \emph{tail} and
its \emph{head}, respectively.  Figure \ref{fr-directed-hypergraphs}
depicts some examples of diercted hypergraphs, where the sets $e_i$ and $e_j$ are
highlighted.

Starting from the definitions of curvature for an arc in the directed
graph case (Equation \ref{in-out-directed-graph}), we introduce the
curvatures $F(_\rightarrow e)$ and $F(e_\rightarrow)$ for a hyperarc as 
\begin{equation}
    \begin{split}
    F(_\rightarrow e)&=|e_i|-\sum_{i\in e_i}\text{in}(i)\\
    F(e_\rightarrow)&=|e_j|-\sum_{j\in e_j}\text{out}(j)
    \end{split}
    \label{flow-curvature-hypergraphs-directed}
\end{equation}
with bounds $|e_i|(1-|E|)\leq F(_\rightarrow e)\leq |e_i|$ and
$|e_j|(1-|E|)\leq F(e_\rightarrow)\leq |e_j|$. For the simple directed
hypergraphs, we have $|e_i|(1-2^{|V|-1})\leq F(_\rightarrow e)\leq |e_i|$
and $|e_j|(1-2^{|V|-1})\leq F(e_\rightarrow)\leq |e_j|$. With
$F(_\rightarrow e)$ and $F(e_\rightarrow)$ at hand, we define the curvature
for the flow through $e=(e_i,e_j)$ as:
\begin{equation}
    \begin{split}
    F(_\rightarrow e_\rightarrow)&=F(_\rightarrow e)+F(e_\rightarrow)\\
     &=|e_i|+|e_j|-\sum_{i\in e_i}\text{in}(i)-\sum_{j\in e_j}\text{out}(j)
    \end{split}
    \label{total-curvature-hypergraphs-directed}
\end{equation}
with bounds
$(1-|E|)(|e_i|+|e_j|)\leq F(_\rightarrow e_\rightarrow) \leq |e_i|+|e_j|$
in the general case and
$(1-2^{|V|})(|e_i|+|e_j|)\leq F(_\rightarrow e_\rightarrow) \leq
|e_i|+|e_j|$ for the simple directed hypergraph (Figure
\ref{fr-directed-hypergraphs}).  Note that if $|e|$ is allowed to have its
minimum value of 1, then $|e_k|=1$ and
$2(1-|E|)\leq F(_\rightarrow e_\rightarrow)\leq 2$.  Some examples of
curvature values for directed hypergraphs are shown in Figure
\ref{fr-directed-hypergraphs}.

The respective flow-loss curvatures are:
\begin{equation}
    \begin{split}
    F(_\leftarrow e)&=|e_i|-\sum_{i\in e_i}\text{out}(i)\\
    F(e_\leftarrow)&=|e_j|-\sum_{j\in e_j}\text{in}(j)
    \end{split}
    \label{contra-curvature-hypergraphs-directed}
\end{equation}
with bounds $|e_i|(1-|E|)\leq F(_\leftarrow e)\leq 0$ and
$|e_j|(1-|E|)\leq F(e_\leftarrow)\leq 0$ in the general case and
$|e_i|(1-2^{|V|})\leq F(_\leftarrow e)\leq 0$ and
$|e_j|(1-2^{|V|})\leq F(e_\leftarrow)\leq 0$ for the simple directed
hypergraphs.

Equation \ref{contra-curvature-hypergraphs-directed} yields the flow-loss
curvature
\begin{equation}
    \begin{split}
    F(_\leftarrow e_\leftarrow)&=F(_\leftarrow e)+F(e_\leftarrow)\\
     &=|e_i|+|e_j|-\sum_{i\in e_i}\text{out}(i)-\sum_{j\in e_j}\text{in}(j)
    \end{split}
    \label{total-contra-curvature-hypergraphs-directed}
\end{equation}
with bounds $(1-|E|)(|e_i|+|e_j|)\leq F(_\leftarrow e_\leftarrow)\leq 0$,
which becomes
$(1-2^{|V|-1})(|e_i|+|e_j|)\leq F(_\leftarrow e_\leftarrow)\leq 0$ for
simple directed hypergraphs.

In the following section we calculate the Forman-Ricci curvature for
different cases that can be modelled as hypergraphs.  Several applications
of the Forman-Ricci curvature for the graph case are found in references
\cite{2017arXiv171207600S, 2017arXiv170700180W, 74177, 2017CSF10150S,
  2016arXiv160708654W, 2016arXiv160807838W, 2016arXiv160504662S,
  2016arXiv160406634W, 2016JSMTE063206S, Jost-MATCH}.

\section*{Applications to Empirical Networks}

\subsection{Wikipedia Voting Network}





Wikipedia is an encyclopedia written by volunteers. A small part of
these users are administrators, who besides being active, regular long-term
Wikipedia contributors, have gained the general trust of the community and
have taken on technical maintenance duties. A user becomes an administrator
when a request for adminship is issued and the Wikipedia community via a
public vote decides who to promote to administrator.  Users can either
submit their own requests for adminship or may be nominated by other
users. Using the January 3 2008 dump of Wikipedia page edit history
\cite{konect:2017:elec}, Leskovec et al. \cite{konect:leskovec207}
extracted 2,794 elections (hyperedges in our setting) and 7,066 users
(vertices) participating in the elections (either casting a vote or being
voted on). We calculated the curvature for the resulting undirected
hypergraph.

\begin{figure}[htb]
    \includegraphics[width=.5\textwidth]{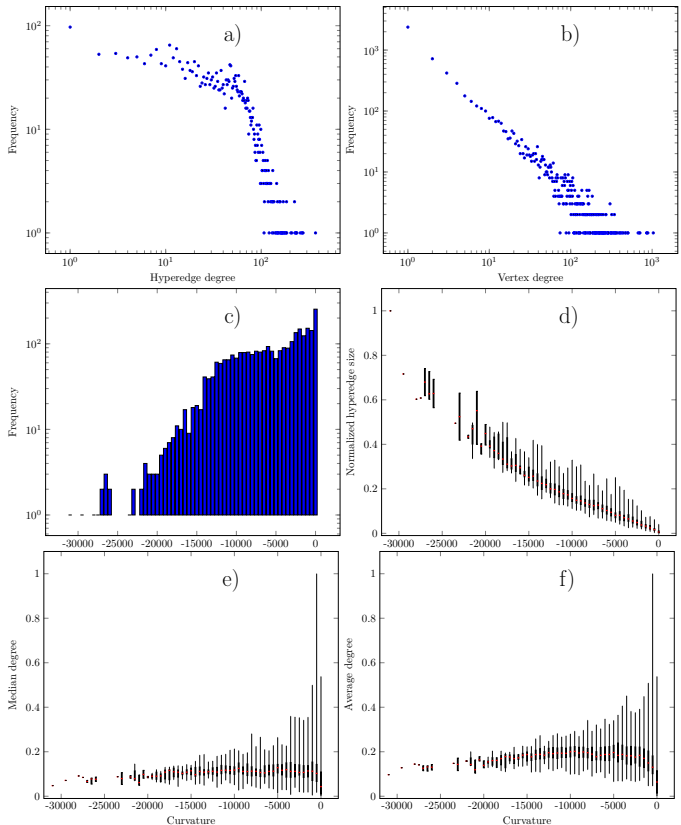}
    \caption{Voting Wikipedia: Distribution of a) hyperedge size (size of
      elections) and b) vertex degree (participation of users in
      elections). c) Histogram of curvature values with bins of 10 units.
      Box-plots of d) normalized hyperedge sizes, e) median, and f) average
      hyperedge degrees corresponding to each curvature bin of c).}
    \label{wiki_elections_deg}
\end{figure}

Figure \ref{wiki_elections_deg} shows the distribution of hyperedge size
and of vertex degree. The data show that many of the elections involve a
single user, although elections with 2-20 users are also common.  There are
few elections with more than 100 users, the largest one including 370 users
(Figure \ref{wiki_elections_deg}a).  The participation in elections
{is} heavy-tailed distributed (Figure \ref{wiki_elections_deg}b),
with most of the users participating in a single election and very few
taking part in about a thousand elections.  The curvature values are mostly
negative (Figure \ref{wiki_elections_deg}c), indicating (i) the absence of
elections with unexperienced users ($\max F(e)\neq |e|$), i.e., all
elections at least include a user that takes part in at least one other
election; and (ii) for most elections the number of elections in which
users take part is greater than their number of voting users ($D>|e|$ in
Equation \ref{curvature-sgD}).  The minimum curvature value (-3,112) is far
from the lower bound ($-19,728,272$, calculated with
$|e|=7,066$). This reflects the fact that most users are experts in
  limited fields only.

To have some insight about the effects of hyperedge sizes and number of
incident hyperedges on curvature, we analyzed their distributions over the
span of curvature values (Figures \ref{wiki_elections_deg}d-f).  Figure
\ref{wiki_elections_deg}d shows that the more spread the election,
i.e. involving users that vote in other elections, the larger the number of
users voting.  Figures \ref{wiki_elections_deg}e and f show that, in
average, elections overlap with a low number of other elections (low number
of incident elections).  Thus, the curvature values are mainly ruled by
hyperedge size rather than by incident hyperedges.


\subsection{Metabolic Network of \textit{Escherichia coli}}

The metabolism of \emph{Escherichia coli} is one of the most studied and
best characterized among bacteria.  Here we model the metabolism K-12 (iJR904
GSM/GPR) \cite{Reed2003} of this bacterium as a directed hypergraph whose
vertices are the metabolites (chemical species). Each chemical reaction is
represented as a hyperarc $e$, whose educts (starting materials) correspond
to $e_i$ and products to $e_j$.  There are $|V|=625$ metabolites and
$|E|=1,176$ reactions accounting for 686 non-reversible and 245 reversible ones.  These latter reactions, denoted by $e_i\leftrightarrow e_j$ have been included as ``forward'' ($e_i\rightarrow e_j$) and ``backward'' ($e_j\rightarrow e_i$) reactions.  All curvatures (Equations \ref{flow-curvature-hypergraphs-directed} to
\ref{total-contra-curvature-hypergraphs-directed}) and related calculations
are gathered in the Supplementary Material.

\begin{figure}[h]                                             
  \centering
  \includegraphics[width=.45\textwidth]{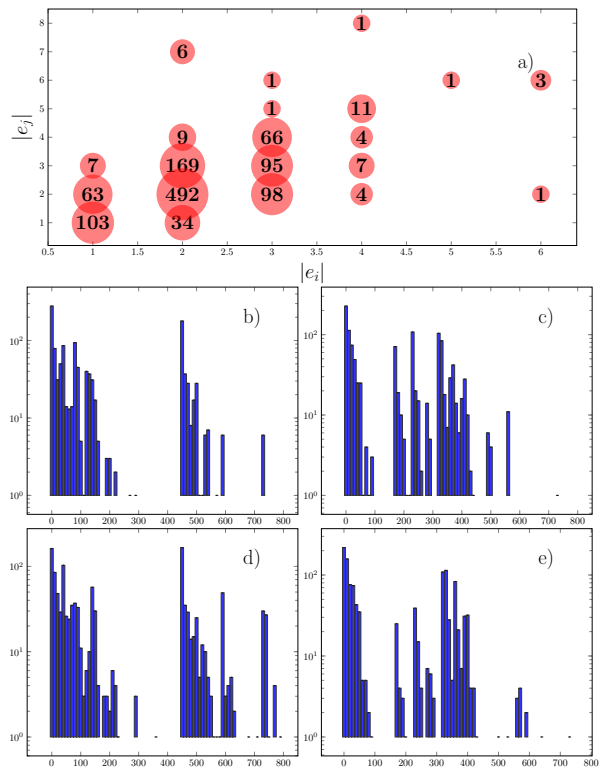}
  \caption{Metabolic network: a) Scatter plot of sizes of educts ($|e_i|$)
    and products ($|e_j|$), where circle radii correspond to
    $\log f/\log 100$, being $f$ the frequency of appearance of the couple
    $(|e_i|,|e_j|)$ in the reactions.  Numbers inside circles correspond to
    $f$. Distribution of b) $\sum_{i \in e_i}{\text{in}(i)}$, c)
    $\sum_{i\in e_i}{\text{out}(i)}$, d) $\sum_{j\in e_j}{\text{in}(j)}$,
    and $\sum_{j\in e_j}{\text{out}(j)}$.}
\label{degrees-metabolism}
\end{figure}

As expected for chemical reactions, typically there are not more than
three educts and three products (Figure \ref{degrees-metabolism}a). The
curvature values therefore vary little in response to hyperarc size, but
rather depend more on the degree of vertices in $e_i$ and $e_j$.  Note that
these degrees result, respectively, from the summation over vertex degrees
of educts and of products (Equations
\ref{flow-curvature-hypergraphs-directed} to
\ref{total-contra-curvature-hypergraphs-directed}).  The distribution and
educts and products degrees is shown in Figure \ref{degrees-metabolism}.
The participation of educts and products in reactions does not yield a
  smooth distribution, as indicated by the gaps present in Figure
  \ref{degrees-metabolism}b,c.  The production of educts (Figure
\ref{degrees-metabolism}b) shows a large group of reactions whose educts
are synthesized by less than 200 reactions and another group where they are
obtained by more than 450 reactions.  Likewise, there are two groups of
reactions with different levels of use of their products (Figure
\ref{degrees-metabolism}c); one group has reactions whose products are used
in less than 100 reactions and another with more than 300 reactions taking
their products as starting materials.

The synthesis of products and the use of educts (Figures
\ref{degrees-metabolism}d and e), shows also a discontinuous participation
of substrates in reactions.  There are two groups of reactions according to
the number of reactions synthesizing their products: one with reactions
whose products are obtained by less than 200 reactions and another by more
than 450 reactions (Figure \ref{degrees-metabolism}d).  Likewise, there are
various groups of reactions according to the use of their educts, from some
which are seldom used to some others with about 170, 230, and more than 330
uses \ (Figure \ref{degrees-metabolism}e).

\begin{figure}[h]                                             
  \centering \includegraphics[width=.45\textwidth]{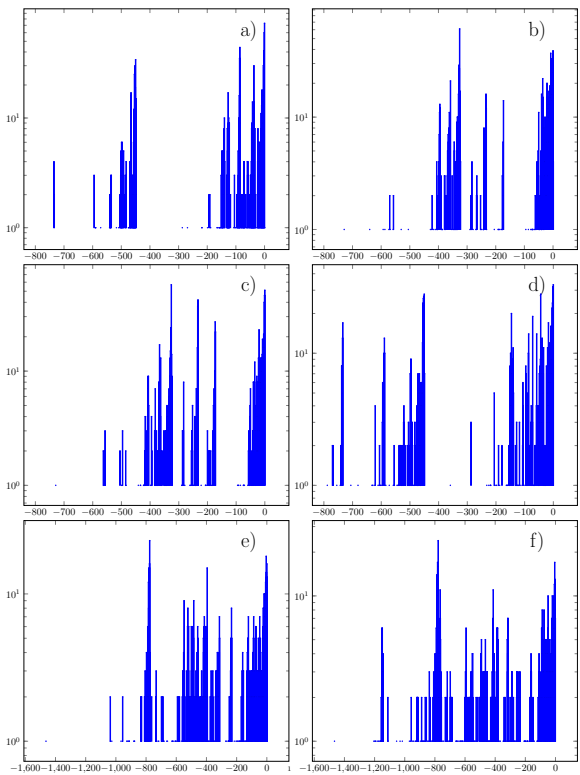}
  \caption{Metabolic network: Histograms of curvature values for a)
    $F(_\rightarrow e)$, b) $F(e_\rightarrow)$, c) $F(_\leftarrow e)$, d)
    $F(e_\leftarrow)$, e) $F(_\rightarrow e_\rightarrow)$, and f)
    $F(_\leftarrow e_\leftarrow)$ with bins of 500 units.}
\label{fr-metabolism}
\end{figure}

The extent to which the educts of the reaction $e$ are produced from
  other reactions is measured by $F(_\rightarrow e)$. The more reactions
lead to the educts of $e$, the more negative $F(_\rightarrow e)$ becomes
(Figure \ref{fr-metabolism}a). The theoretical bounds of
$F(_\rightarrow e)$, assuming $\max|e_i|=625$ are
$-734,375 \leq F(_\rightarrow e)\leq 625$. However, more realistic bounds
are $-7,050 \leq F(_\rightarrow e)\leq 6$, which results from taking the
actual $\max|e_i|=6$ (Figure \ref{degrees-metabolism}a).  We found that
$\min F(_\rightarrow e)=-735$, which is attained by four reactions, with
four educts (all substrate abbreviations are included in the Appendix
(Table \ref{substrates})):

\par\noindent
\chem{adp}+\chem{h}+\chem{malcoa}+\chem{pi}$\rightarrow$\chem{accoa}+\chem{atp}+\chem{hco3}\\
\chem{adp}+\chem{h}+\chem{pi}+\chem{25aics}$\rightarrow$\chem{asp\_L}+\chem{atp}+\chem{5aizc}\\
\chem{adp}+\chem{dtbt}+\chem{h}+\chem{pi}$\rightarrow$\chem{atp}+\chem{co2}+\chem{dann}\\
\chem{atp}+\chem{gar}+\chem{h}+\chem{pi}$\rightarrow$\chem{atp}+\chem{gly}+\chem{pram}
\par\noindent

These reactions are those whose educts are the most synthesized of all the
metabolic reactions of \emph{E. coli} (63\% of the reactions produce their
educts).  In three of them \chem{atp} is synthesized from \chem{adp}, which shows the well-known central metabolic role of \chem{atp} \cite{Albert2000, Wagner1803}.

$\max F(_\rightarrow e)=1$ corresponds to a single reaction:
\chem{cyan}+\chem{tsul}$\rightarrow$\chem{h}+\chem{so3}+\chem{tcynt}, where only one of its two educts is a product of a single reaction: \chem{atp}+\chem{h2o}+\chem{tsul}$\rightarrow$\chem{adp}+\chem{h}+\chem{pi}+\chem{tsul}.

Figure \ref{fr-metabolism}a shows that the most frequent curvature value is
0 (for 73 reactions), i.e. 6\% of the reactions have a trade-off between
the number of educts and the number of reactions producing them; most of
the remaining reactions have more ways to produce their educts than the
number of educts.  It is also found that there are almost no reactions with
curvatures between -200 and -450, indicating that educts of reactions are
mainly obtained either by less than 200 reactions (less than 17\% of the
reactions) or by 450 to 600 reactions (38 to 51\% of the reactions).  This
is a consequence of the heavy-tailed in-degree distribution of substrates
\cite{Albert2000}.

Figure \ref{fr-metabolism}b shows the curvature values $F(e_\rightarrow)$,
which quantify the extent to which products of reactions are used in
further reactions as educts.  By taking $\max |e_j|=8$ (Figure
\ref{degrees-metabolism}a) this curvature takes values
$-9,400\leq F(e_\rightarrow)\leq 8$.  The actual
$\min F(e_\rightarrow)=-729$, for
\chem{adp}+\chem{h}+\chem{pi}$\rightarrow$\chem{atp}+\chem{h}+\chem{h2o},
i.e., this is the reaction whose three products are most used in other
reactions as starting materials (used in 62\% of the reactions).  In
contrast, there are four reactions with $\max F(e_\rightarrow)=1$:
\par\noindent
\chem{agpe\_EC} + \chem{pg\_EC} $\rightarrow$ \chem{apg\_EC}+\chem{g3pe}\\
\chem{agpc\_EC} + \chem{pg\_EC} $\rightarrow$ \chem{apg\_EC}+\chem{g3pc}\\
\chem{agpg\_EC} + \chem{pg\_EC} $\rightarrow$ \chem{apg\_EC}+\chem{g3pg}\\
\chem{udpgal}   $\rightarrow$ \chem{udpgalfur}
\par\noindent
Hence, for those three reactions with two products, these substrates are
only used in a further reaction as educts, while \chem{udpgalfur} is not further
used, i.e. it is a metabolic ``dead-end'' \cite{Reed2003}.  As most of the
reactions (96\%) have negative values of $F(e_\rightarrow)$, this indicates
the efficient use of reaction products \cite{Wagner1803}, which can be
divided into two regimes.  For about half of the reactions their products
are used in no more than 9\% of the reactions and about 40\% of the
reactions have products that are used in more than a quarter of the
reactions.  This is a consequence of the heavy-tailed distribution, this
time, of the out-degrees of the substrates \cite{Albert2000}.

$F(_\rightarrow e)$ showed that for most of the reactions their educts are
produced by other reactions and $F(e_\rightarrow)$ that the products are
used in several other reactions.  The question that arises whether those
popular educts are connected through reactions with the popular products is
positively answered by $F(_\rightarrow e_\rightarrow)$, which takes
negative values for most of the reactions.  The
$\min F(_\rightarrow e_\rightarrow)=-1,463$ corresponds to
\chem{adp}+\chem{h}+\chem{pi}$\rightarrow$\chem{atp}+\chem{h}+\chem{h2o}.
Hence, this is the reaction whose educts are most synthesized by other
reactions and whose products are the most used as educts in other
reactions.  It is the bottleneck of the \emph{E.\ coli} metabolism.  Other
reactions of this sort, with $F(_\rightarrow e_\rightarrow)<-1,000$ (Figure
\ref{fr-metabolism}e), are:
\par\noindent
\chem{adp}+\chem{h}+\chem{pi}$\rightarrow$\chem{atp}+\chem{h}+\chem{h2o}\\
\chem{h}+\chem{o2}+\chem{q8h2}$\rightarrow$\chem{h}+\chem{h2o}+\chem{q8}\\
\chem{h}+\chem{o2}+\chem{q8h2}$\rightarrow$\chem{h}+\chem{h2o}+\chem{q8}\\
\chem{h}+\chem{no3}+\chem{q8h2}$\rightarrow$\chem{h}+\chem{h2o}+\chem{no2}+\chem{q8}\\
\chem{h}+\chem{mql8}+\chem{no3}$\rightarrow$\chem{h}+\chem{h2o}+\chem{mqn8}+\chem{no2}
\par
Having analyzed the metabolism following the direction of educts to
products in reactions, we now proceed to study the curvature in the
backward direction, which quantifies to which extent a reaction is just one
of the many connecting popular educts with popular products.  We start by
analyzing $F(_\leftarrow e)$ that shows to which extent educts of a
reaction participate in other reactions.  The theoretical bounds are
$-734,375\leq F(_\leftarrow e)\leq 0$ and we found that $F(_\leftarrow e)$
takes values in between -729 and 0; the minimum is attained by
\chem{atp}+\chem{h}+\chem{h2o}$\rightarrow$\chem{adp}+\chem{h}+\chem{pi}, indicating that \chem{atp} in an acidic aqueous
medium is the most often used starting material.  $\max F(_\leftarrow e)$
occurs for 51 reactions, whose involved 56 educts are only used in those 51
reactions, i.e. they are very specialized educts for very particular
metabolic reactions.  The distribution of $F(_\leftarrow e)$ values shows
that for half of the reactions, their educts participate in less than 9\%
of the reactions, while for the rest, their educts take part in more than
15\% of the reactions.


$F(e_\leftarrow)$ shows to which extent products of a reaction are
synthesized by other reactions.  The theoretical bounds are given by
$\max |e_j|=8$, leading to $-9,400\leq F(e_\leftarrow) \leq 0$.  The actual
values range from -788 to 0.  The minimum is reached by reaction:
\chem{dxyl5p}+\chem{nad}+\chem{phthr}$\rightarrow$\chem{co2}+\chem{h}+\chem{h2o}+\chem{nadh}+\chem{pdx5p}+\chem{pi}, i.e. this set of
products is the most synthesized by \emph{E. coli} metabolism, which is
expected, for the likelihood of a set of substances to be synthesized
scales with the size of the set.  This reaction with six products is one of
the few where more than the frequent one to four products are synthesized
(Figure \ref{degrees-metabolism}a).  Moreover, among the products,
\chem{co2}, \chem{h}, \chem{h2o}, \chem{nadh}, and \chem{pi} are often products of other
reactions.

$\max F(e_\leftarrow)=0$ is attained by 29 reactions, all of them leading
to a single product, except for three reactions, each one with two
products.  Thus, those 32 products are of little synthetic relevance for
the metabolism.  The distribution of curvature values shows that there are
three kinds of reactions whose products are synthesized by different number
of reactions.  For 60\% of the reactions their products are synthesized by
less than 200 reactions (17\% of the reactions) and for the rest of the
reactions by more than 450 reactions (38\% of the reactions).


Curvatures $F(_\leftarrow e)$ and $F(e_\leftarrow)$ showed that half of the
educts are often used and 40\% of the products are often synthesized, which
indicates that it is very likely to find alternative ways to link educts
with products of existing reactions, as found in \cite{Albert2000,
  Barabasi509, Wagner1803}.  A measure of this degree of redundancy of a
reaction or of its replaceability is given by
$F(_\leftarrow e_\leftarrow)$, which indicates to which extent a reaction
connects popular educts with popular products.  The more negative the
curvature, the more redundant or likely replaceable the reaction is.

By analyzing $F(_\leftarrow e_\leftarrow)$
  distribution (Figure \ref{fr-metabolism}f) it is seen an ample spectrum
  of curvatures, with almost no gaps, indicating different degrees of
  redundancy for the metabolic reactions.  $\min F(_\leftarrow e_\leftarrow)=-1,463$ corresponds to
the hydrolysis of ATP, i.e.,
\chem{atp}+\chem{h}+\chem{h2o}$\rightarrow$\chem{adp}+\chem{h}+\chem{pi},
indicating, e.g., that the dephosphorilation of \chem{atp} to \chem{adp} can be achieved
by many other reactions (12\% of the reactions).
$\max F(_\leftarrow e_\leftarrow)=0$ occurs for the following eight
reactions, which are unique as they are the only way to connect their
educts with their products:
\par\noindent
\chem{mmcoa\_R}$\rightarrow$\chem{mmcoa\_S}\\
\chem{5mdr1p}$\rightarrow$\chem{5mdru1p}\\
\chem{gdpddman}$\rightarrow$\chem{gdpofuc}\\
\chem{adphep\_D,D}$\rightarrow$\chem{adphep\_L,D}\\
\chem{dhnpt}$\rightarrow$\chem{gcald}+\chem{6hmhpt}\\
\chem{glu1sa}$\rightarrow$\chem{5aop}\\
\chem{prfp}$\rightarrow$\chem{prlp}\\
\chem{pran}$\rightarrow$\chem{2cpr5p}
\par\noindent

\section*{Conclusions and Outlook}

The Forman-Ricci curvature emphasizes the importance of the relational
  character of (hyper)edges, thereby providing a view of the network
  structure that complements traditional vertex-centered descriptors. It
  also embeds the characterization in a formal mathematical theory, namely
  Riemannian geometry.

The results here reported include a brief review of Forman-Ricci curvature
for (un)directed graphs and generalize the curvature to both undirected and
directed hypergraphs.  Graph curvatures used in previous studies thus become particular cases of the curvature
for hypergraphs \cite{2017arXiv171207600S, 2017arXiv170700180W, 74177, 2017CSF10150S,
    2016arXiv160708654W, 2016arXiv160807838W, 2016arXiv160504662S,
    2016arXiv160406634W, 2016JSMTE063206S, Jost-MATCH}.  We determined the upper and lower bounds for Forman-Ricci
curvature for graphs and hypergraphs, which so far had not been studied.

The curvatures here presented aim at quantifying the trade-off between
hyperedge(arc) size and the degree of participation of vertices members of
the hyperedge(arc) in other hyperedges(arcs).  For undirected hypergraphs,
the curvature takes negative values when the degree of vertices of the
hyperedge is more significant than the size of the hyperedge.  For directed
hypergraphs we devised four curvatures that gauge different aspects of
hyperarcs.  $F(e_\rightarrow)$ quantifies the trade-off between the size of the 
hyperarc tail and the input of its vertices, $F(_\rightarrow e)$ do so for
the size of the hyperarc head and the output of its vertices; while
$F(_\leftarrow e)$ and $F(e_\leftarrow)$ consider the size of the tail and
the output of their vertices, and the size of the head and the input of
their vertices, respectively.  These curvatures are combined into
$F(_\rightarrow e_\rightarrow)$ and $F(_\leftarrow e_\leftarrow)$, which
account for the flow through hyperarc $e$ and for its redundancy,
respectively.
 
The Forman-Ricci curvature for hypergraphs introduced here differs
  from the alternative construction proposed in \cite{Emil-Melanie}. There,
  hyperedges are interpreted as simplices. Here, we focus entirely on
  hyperedges, their sizes, and the degrees of vertices, thereby avoiding
  the re-interpretation of hyperedges as ``higher-dimensional'' objects
 and implicitly introducing additional structures, like boundaries of simplices, that  are not part of
  the original data.  For the particular case of directed
hypergraphs we disentangled the curvature in the four aforementioned
informative measures that allow a detailed exploration of the hypergraph
structure.  Moreover, we applied these curvatures to the analysis of two
large networks, one of social and the other of chemical interactions.

The analysis of Wikipedia vote network exemplified the Forman-Ricci
curvature of undirected hypergraphs, where elections constituted hyperedges
and users/voters vertices.  We found that curvature is mostly ruled by
hyperedge size rather than by hyperedge degree.  Likewise, the more users
involved in elections, the more the presence of experienced users.  In a
traditional graph setting \cite{konect:2017:elec, konect:leskovec207}, with
users as vertices and votes as couple of users, conclusions such as the
previous one on elections cannot be drawn.  This shows the richness of
hypergraphs and their curvatures, which for this particular case allowed
the definition of a hyperedge as an election.

Forman-Ricci curvature for directed hypergraphs was computed over the
metabolic network of \emph{E. coli}, which traditionally has been analyzed
through a graph setting \cite{Albert2000, Barabasi509, Wagner1803} and
which has shed light on the important role of several substrates for the
metabolic stability.  In our approach, rather than focusing on substrates,
we did on reactions, which were characterized as hyperarcs connecting sets
of educts with sets of products.

In contrast to the Wikipedia vote network, we found that curvature values
for the metabolic network were ruled by the degree of hyperarcs, i.e. of
in- and out-degrees of tails and heads of hyperarcs.  This is a chemical
consequence, for it is unlikely that several educts collide simultaneously
to give place to a product.  In fact reactions where more than five educts
participate in a single-step reaction are scarce \cite{C3CS35505E}.

We emphasize that the strong dependence of hyperarc curvature is on the
summation of the degrees of the vertex belonging to the hyperarc, which is
different from the traditional degree of isolated vertices.

With curvature results at hand we defined ``bottle neck'' reactions as
those few reactions whose educts are readily available (obtained from
several reactions) and whose products are often used as educts.  They are
characterized by having very negative $F(_\rightarrow e_\rightarrow)$
values.  For \emph{E.\ coli} this reactions is:
\chem{adp}+\chem{h}+\chem{pi}$\rightarrow$\chem{atp}+\chem{h}+\chem{h2o}.
Bottle neck reactions can be considered as assortative ones, for they
transform popular products into popular educts.

Curvature values also allowed detecting redundant reactions (``one of the
crow reactions''), which can be easily replaced by others.  The suitable
curvature for detecting such reactions is $F(_\leftarrow e_\leftarrow)$,
whose most negative values correspond to reactions where popular sets of
educts are connected to popular sets of products.  For \emph{E. coli}, this
reaction is
\chem{atp}+\chem{h}+\chem{h2o}$\rightarrow$\chem{adp}+\chem{h}+\chem{pi}.
Thus, \chem{adp} phosphorilation is the metabolic bottle neck reaction but
the reverse reaction is not that central for the metabolism.

Our results show that \emph{E.\ coli} metabolic network makes use of a
wealth amount of the products of its reactions to start other reactions.
This contrast with the historical trend in wet-lab chemistry reactions,
where most of the products are seldom used in further reactions
\cite{PNAS}.  As the historical study was conducted over single substances,
rather than over educts and products, further work on the curvature of
wet-lab chemical reactions needs to be done to determine whether the
behaviour found for \emph{E.\ coli} is also a trend of chemical reactions,
in general.

The curvatures here presented, as indicated in Equation \ref{FR-curvature}
and as used in \cite{2017arXiv171207600S, 2017arXiv170700180W, 74177,
  2017CSF10150S, 2016arXiv160708654W, 2016arXiv160807838W,
  2016arXiv160504662S, 2016arXiv160406634W, 2016JSMTE063206S, Jost-MATCH},
can be weighted.  In the recent sketch of curvature for hypergraphs
\cite{Emil-Melanie}, the weights are calculated based on the volume of the
simplex associated to the hyperedge.  Weights, however, can also be based
on meta information of the network, e.g.\ user's seniority in the Wikipedia
example or stoichiometric coefficients in the metabolic network.  This and
other weighting schemes need to be explored in future studies on the
curvature of hypergraphs, which our approach allows.


\section*{Acknowledgments}
WL was supported by a PhD scholarship from the \emph{German Academic Exchange
    Service} (DAAD): Forschungsstipendien-Promotionen in Deutschland, 2017/2018 (Bewerbung 57299294).

\section{Appendix}

\begin{table}[h]
\centering

\label{substrates}
\end{table}

\bibliography{ms} 
\bibliographystyle{ieeetr}

\end{document}


\section*{Supplementary Material}
Curvature values for chemical reactions from the metabolism of \emph{Escherichia coli} K-12 (iJR904 GSM/GPR).
    %
